\documentclass[pdflatex, sn-mathphys-num, 10pt]{sn-jnl}

\usepackage{mathptmx}

\usepackage{graphicx}%
\usepackage{multirow}%
\usepackage{amsmath,amssymb,amsfonts}
\usepackage{bm}       

\usepackage{amsthm}%
\usepackage{mathrsfs}%
\usepackage[title]{appendix}%
\usepackage{xcolor}%
\usepackage{textcomp}%
\usepackage{manyfoot}%
\usepackage{booktabs}%
\usepackage{algorithm}%
\usepackage{algorithmicx}%
\usepackage{algpseudocode}%
\usepackage{listings}%

\usepackage{subcaption}

\theoremstyle{thmstyleone}%
\theoremstyle{thmstyletwo}%

\theoremstyle{thmstylethree}%

\begin{document}

\title[Article Title]{Contextual Checkerboard Denoise - A Novel Neural Network-Based Approach for Classification-Aware OCT Image Denoising}


\author[1]{\fnm{Md. Touhidul} \sur{Islam}}\email{1806177@eee.buet.ac.bd}

\author[1]{\fnm{Md. Abtahi M.} \sur{Chowdhury}}\email{1806106@eee.buet.ac.bd}

\author[1]{\fnm{Sumaiya} \sur{Salekin}}\email{1806191@eee.buet.ac.bd}

\author[1]{\fnm{Aye T.} \sur{Maung}}\email{1806195@eee.buet.ac.bd}

\author[1]{\fnm{Akil A.} \sur{Taki}}\email{1806139@eee.buet.ac.bd}

\author*[1]{\fnm{Hafiz} \sur{Imtiaz}}\email{hafizimtiaz@eee.buet.ac.bd}

\affil[1]{\orgdiv{Department of Electrical and Electronic Engineering}, \orgname{Bangladesh University of Engineering and Technology}, \orgaddress{\city{Dhaka}, \country{Bangladesh}}}

\abstract{In contrast to non-medical image denoising, where enhancing image clarity is the primary goal, medical image denoising warrants preservation of crucial features without introduction of new artifacts. However, many denoising methods that improve the clarity of the image, inadvertently alter critical information of the denoised images, potentially compromising classification performance and diagnostic quality. Additionally, supervised denoising methods are not very practical in medical image domain, since a \emph{ground truth} denoised version of a noisy medical image is often extremely challenging to obtain. In this paper, we tackle both of these problems by introducing a novel neural network based method -- \emph{Contextual Checkerboard Denoising}, that can learn denoising from only a dataset of noisy images, while preserving crucial anatomical details necessary for image classification/analysis. We perform our experimentation on real Optical Coherence Tomography (OCT) images, and empirically demonstrate that our proposed method significantly improves image quality, providing clearer and more detailed OCT images, while enhancing diagnostic accuracy.}




\maketitle

\section{Introduction}\label{sec1}
Denoising is the task of estimating the original (noise-free) image $\boldsymbol{S}$ from a noisy observation $\boldsymbol{X = S + N}$, where $\boldsymbol{N}$ is the noise signal. In particular, the primary goal of non-medical image denoising is enhancing the clarity of an image. However, for medical image denoising, the major objective is enhancing image clarity \emph{without} i) introducing any new artifacts into the image; ii) altering any essential information in the image. While many of the existing denoising algorithm visibly improves image clarity by a good margin, the classification and diagnostic performance of these denoised images rather under-perform due to the trade-off of image originality for image clarity. As such, it is crucial that the denoising of the medical images is performed in a way that it complements future classification, analysis and/or diagnosis process. We note that, a practical limitation for medical image denoising is that it is difficult to obtain ground truth denoised versions of noisy images for supervised learning training purposes. Self-supervised medical image denoising models are, therefore, useful in numerous medical applications.

Optical Coherence Tomography (OCT) images are good examples of medical imagery, where noisy images pose frequent problems. OCT is a non-invasive imaging modality that has become increasingly important in clinical diagnostics, particularly in ophthalmology and cardiology. OCT images are often degraded by speckle noise and limited sampling rates, which can obscure critical details and hinder accurate interpretation. The coherent nature of the light source also leads to the formation of speckle noise, which can reduce contrast and obscure important features. Additionally, practical limitations on the scanning speed and detector size often result in OCT images with relatively low sampling rates, which can lead to grainy and aliased appearances.

Recent advances in OCT image denoising leverage a variety of deep learning approaches. Dong et al.~\cite{dong2020optical} used a generative adversarial network (GAN) with speckle modulation for speckle noise reduction. However, GAN-based architectures inherently tend to \emph{generate} new information in the image. This can potentially lead to a poor quality with respect to medical applications and image classification results after denoising. Bayhaqi et al.~\cite{9760477} introduced a CNN that is optimized for single-image efficiency, while simulating frame-averaging. Jose et al.~\cite{Rico} introduced a self-fusion network designed to exploit similarities between adjacent scans for real-time denoising without extensive training data. Although these techniques are optimized to produce clean and denoised images, they rely on ground truth denoised images for supervised denoising, which is impractical for most medical image denoising tasks. Innovative frameworks, such as Neighbor2Neighbor~\cite{huang2021neighbor2neighbor}, Noise2Void~\cite{n2v}, and PN2V~\cite{pn2v} have introduced self-supervised denoising methods that avoid the need for noise modeling or clean image pairs. These techniques exploit spatial correlations within noisy images to train neural networks, effectively leveraging local patterns without relying on ground truth data. However, these methods are not context- or classification-aware. They treat the image as a collection of independent pixels, ignoring any higher-level semantic structure or contextual information that might be crucial for further image analysis/classification. This lack of contextual awareness can lead to the loss of important image details or misclassifications, especially in scenarios where global or class-specific information plays a vital role in denoising effectiveness.

To address the aforementioned research gap, we propose a novel neural network architecture that takes into account the key requirements of effective medical image denoising -- i) cannot introduce any new artifact into the denoised image; ii) cannot remove essential clinical detail of the image; iii) should be able to denoise from single noisy images dataset. We introduce a custom loss function for training our neural network that takes both classification loss and denoising loss into account. As a result, our training process simultaneously enhances the image and complements the diagnosis/classification performance of the image. Additionally, we introduce a novel blind spotting technique for self-supervised denoising, which is simple and efficient, and very effective on noisy OCT image dataset.

\section{Related Works}
Image denoising is a crucial pre-processing step in medical image analysis, as noise impairs the clarity of the image, impacting the diagnostic accuracy. Unlike non-medical images, preserving key features, such as edges and textures, is necessary in denoising OCT B-scans. The numerous methods applied to reduce speckle noise utilize basic filtering, wavelet-based techniques, model-based approaches, and modern methods leverage deep learning and hybrid techniques.

Traditional denoising methods like Gaussian filtering, median filtering, and mean filtering are commonly used due to their simplicity and efficiency. Gaussian filtering reduces noise by averaging pixel values with a Gaussian-weighted sum, but it can blur fine details and edges~\cite{mallat1999wavelet}. Median filtering, on the other hand, is more effective in removing salt-and-pepper noise by selecting the median value within a pixel neighborhood, preserving edges better than Gaussian filtering~\cite{mallat1999wavelet}. Mean filtering, as a straightforward averaging technique, smooths images but suffers from similar limitations in edge preservation.

Wavelet-based denoising leverages the multi-resolution nature of wavelets to separate noise from important image features. Techniques such as BM3D (Block-Matching and 3D filtering) utilize a 3D transformation by grouping similar patches and applying collaborative filtering in a transformed domain. This approach effectively reduces noise in medical images while preserving details~\cite{chen2017trainable}. Complex wavelet transforms, like complex wavelet-based K-SVD, offer enhanced noise suppression by decomposing the image into high- and low-frequency components, processing each independently~\cite{chen2017trainable}.

Model-based denoising methods, such as the Wiener filter and general Bayesian estimation, rely on statistical models to estimate and reduce noise. The Wiener filter minimizes the mean square error between the estimated and true signal, often assuming Gaussian noise~\cite{ lim1990two}. Bayesian methods generalize this approach by incorporating prior knowledge about the image or noise, making them adaptable to various noise types encountered in medical imaging~\cite{ lim1990two}. These approaches perform well under specific noise assumptions but may struggle with complex or non-Gaussian noise patterns.

Total Variation (TV) minimization is a regularization method that reduces noise while maintaining edges by minimizing the total variation of the image. It is effective in removing Gaussian noise but can lead to staircasing effects~\cite{chambolle2011introduction}. Total Generalized Variation (TGV) decomposes the image into different regularization terms, further enhancing noise removal and edge preservation. These methods are particularly valuable in medical imaging, where edge sharpness is crucial for accurate analysis~\cite{chambolle2011introduction}.

Non-Local Means (NLM) filtering improves denoising by averaging all pixels in the image with similar local neighborhoods rather than relying only on spatial proximity~\cite{buades2005nonlocal}. This non-local approach is particularly effective in retaining textures and fine details in medical images. However, it can be computationally expensive, limiting its application in high-resolution images.

Hybrid methods like the bilateral filter and MSBTD (Multi-Scale Block Threshold Denoising) combine principles from various denoising techniques. The bilateral filter smooths images while preserving edges by weighting pixel differences based on both spatial proximity and intensity difference~\cite{dabov2007image}. MSBTD employs multi-scale processing to improve denoising at different resolution levels, enhancing its adaptability to complex noise patterns in medical imaging~\cite{fisher2018adaptive}.

With advancements in deep learning, neural network-based approaches are gaining popularity in image denoising. Self-supervised models, such as PN2V (Probabilistic Noise2Void) and Noise2Void, learn to denoise images by only using the noisy image itself as input, eliminating the need for clean training data~\cite{krull2019noise2void}. These models are particularly advantageous for medical images, where acquiring paired noisy-clean datasets can be challenging. These methods have demonstrated superior performance in noise removal and feature preservation compared to traditional techniques, laying the foundation for classification-aware denoising models.\\

\noindent\textbf{Our Contributions.} Despite the progress in denoising methods, current algorithms and models struggle to differentiate between informative and non-informative speckle, leading to potential loss of clinically valuable details. More precisely, there are still issues with efficiently handling intricate and non-Gaussian noise patterns in OCT B-scans while maintaining important features like edges and textures. The need for self-supervised and hybrid deep learning techniques that are adapted to the particular needs of medical imaging is highlighted by this as well as the scarcity of paired noisy-clean datasets. In this paper, we address this research gap by proposing a novel approach, \textit{Contextual Checkerboard Denoise}, which performs model training by combining classification awareness with self-supervised image denoising. Our major contributions are:

\begin{itemize}
    \item We propose a novel neural network-based approach -- Contextual Checkerboard Denoise, which is designed specifically for classification-aware self-supervised medical image denoising.
    \item We propose to integrate contextual information to better preserve clinically relevant features while reducing noise.
    \item We propose a novel blind spotting method that works by alternatively blind spotting even and odd pixels, which shows superior empirical effectiveness.
    \item Finally, we demonstrate the performance of our proposed model in preserving classification-relevant features on noisy OCT images.
\end{itemize}

\noindent\textbf{Notations. }For vector, matrix, and scalar, we used bold lower case letter $(\mathbf{v})$, bold capital letter $(\mathbf{V})$, and unbolded letter ($M$), respectively. We used the symbol $\mathbf{v}_n$ for the $n$-th column of the matrix $\mathbf{V}$; and $v_{ij}$ denotes the $(i,j)$-th entry of matrix $\mathbf{V}$. We sometimes denote the set $\{1, 2, \ldots, N\}$ as $[N]$. Inequality $\mathbf{V} \geq 0$ apply entry-wise. We denoted $\mathcal{L}_2$ norm (Euclidean norm) with $\Vert . \Vert_2$, the $\mathcal{L}_{1,1}$ norm with the $\Vert . \Vert_{1,1}$, and the Frobenius norm with $\Vert . \Vert_F$. $\mathbb{R}$, and $\mathbb{R}_+$ denote the set of real numbers, and set of positive real numbers, respectively.

\section{Background}
In this work, we employ a modified ResUNet++ backbone containing \emph{encoding layers} (i.e., Conv2D, Batch Normalization, ReLU, and SE blocks) that extract features from the noisy image. These layers serve two major purposes:
\begin{itemize}
    \item Denoising the image by learning noise-invariant features.
    \item Extracting meaningful features important for the subsequent classification task.
\end{itemize}
This shared feature extraction ensures that the denoising process is optimized for classification needs. In this section, we briefly discuss the ResUNet++ model architecture, along with some necessary background concepts. Note that, we modified the ResUNet++ architecture by adding a fully connected linear layer at the end of the encoder block. This facilitates the integration of our custom classification-aware loss into the model.

\subsection{ResUNet++ Architecture}
ResUNet++ is an improved ResUNet architecture~\cite{resunet}, leveraging advanced deep learning techniques that includes convolutional blocks, Squeeze and Excite (SE) blocks, Atrous Spatial Pyramid Pooling (ASPP), and attention mechanisms. This architecture outperforms both ResUNet and U-Net deep-learning architectures, as it enhances the denoising performance by jointly optimizing for both denoising and classification loss. Moreover, it effectively removes noise, while preserving critical features necessary for accurate medical diagnosis. The major parts of the ResUNet++ architecture are described in the following.

\subsubsection{Input Layer}
The model takes a noisy image as input, denoted as $\boldsymbol{X}$. As mentioned before, the noisy image is modeled as 
\[
\boldsymbol{X = S + N},
\]
where $\boldsymbol{X}$, $\boldsymbol{S}$, and $\boldsymbol{N}$ are 2D matrices representing the pixel intensities of the image, true noise-free image, and additive noise, respectively. The input is then passed through a series of convolutional operations for feature extraction and denoising.

\subsubsection{Stem Block and Encoder Blocks}
The stem block incorporates the initial stage of the network that extracts features using convolutional layers, followed by batch normalization and ReLU activation to set the groundwork for deeper feature learning~\cite{8470438}. Next, an encoder block contains two successive $3 \times 3$ convolutional layers, combined with batch normalization and ReLU activation. The encoder blocks include an identity mapping that connects the input and output, allowing efficient information flow~\cite{8741187}. A convolution operation then reduces the spatial dimensions to focus on high-level features. 

\subsubsection{Residual Units}
Residual units are used throughout the architecture to address the degradation problem found in deep networks~\cite{lan2021image}. Mathematically, the output $\boldsymbol{y}$ of a residual unit can be expressed as:
\[
\boldsymbol{y} = \boldsymbol{x} + F(\boldsymbol{x}, \boldsymbol{W_i}),
\]
where $\boldsymbol{x}$ is the input, and $F(\boldsymbol{x}, \boldsymbol{W_i})$ represents a series of transformations (e.g., convolutions, batch normalization, and ReLU) parameterized by the weights $W_i$. The skip connections ensure efficient training and better information propagation.

\subsubsection{Squeeze and Excitation (SE) Blocks}
Located after each encoder block, SE blocks enhance feature representation by modeling channel-wise dependencies~\cite{roy2018concurrent}. This involves two main operations:
\begin{itemize}
    \item \textbf{Squeeze}: Global average pooling generates channel-wise statistics.
    \item \textbf{Excitation}: Channel dependencies are modeled to adjust feature relevance. Mathematically, the recalibration is given by:
    \[
    s_c = \sigma \left( \boldsymbol{W_2} \cdot \mathrm{ReLU}(\boldsymbol{W_1} \cdot \boldsymbol{z_c}) \right),
    \]
    where $\boldsymbol{z_c}$ is the squeezed feature vector, $\boldsymbol{W_1}$ and $\boldsymbol{W_2}$ are learnable weight matrices, $\sigma$ denotes the sigmoid function defined as: 
    \[
    \sigma(x) = \frac{1}{1 + e^{-x}}
    \]
    and ReLU denotes the Rectified Linear Unit function defined as:
    \[
    \text{ReLU}(x) = \max(0, x)
    \]
    
\end{itemize}

\subsubsection{Atrous Spatial Pyramid Pooling (ASPP)}
ASPP acts as a bridge between the encoder and decoder, capturing multi-scale contextual information using parallel atrous convolutions with varying rates~\cite{jie2021atrous}. This mechanism expands the receptive field, enabling the model to process and preserve features at multiple scales for enhanced denoising performance, while preventing the removal of critical information needed for classification.

\subsubsection{Decoder Blocks}
The decoding path mirrors the encoder structure and includes residual units. An attention block is placed before each residual unit to prioritize important features and suppress irrelevant ones~\cite{deng2021research}. The decoder uses nearest-neighbor up-sampling to increase spatial resolution, and concatenates these features with those from the encoder to retain fine details.

\subsubsection{Attention Blocks}
Attention blocks enhance the model's ability to focus on crucial regions in the feature maps, improving the denoising performance and reducing computational overhead. The attention mechanism assigns higher weights to relevant features, optimizing the model's focus~\cite{niu2021review}. Mathematically, the attention mechanism can be represented as:
\[
A = \sigma \left( \boldsymbol{W_2} \cdot \mathrm{tanh}(\boldsymbol{W_1} \cdot \boldsymbol{f} + b_1) + b_2 \right),
\]
where $\boldsymbol{f}$ represents the input feature vector, $W_1$ and $W_2$ are learnable weights, and $b_1$ and $b_2$ are biases.

\subsubsection{Output Layer and Loss Functions}
A final $1 \times 1$ convolutional layer with sigmoid activation generates the denoised segmentation map, ensuring that noise is minimized, while retaining critical medical details. To that end, we introduce a novel composite loss function. The losses relevant for our composite loss function are:

\begin{itemize}
    \item \textbf{RMS Loss}: Measures pixel intensity accuracy:
    \[
    \mathrm{RMSLoss} = \frac{1}{N \times N} \sum_{i,j} (v_{ij} - v_{ij}^\prime)^2,
    \]
    where $v$ and $v^\prime$ are the true and predicted pixel values, respectively.
    
    \item \textbf{Cross-Entropy Loss}: Measures the cross-entropy:
    
\[
    CrossEntropyLoss = \sum \log\frac{\exp(\boldsymbol{x'})}{\exp(\boldsymbol{x})}.
\]
Here, $\boldsymbol{x^\prime}$ represents the logit (predicted score) for the correct class, and $\boldsymbol{x}$ is the vector of logits for all classes.
\end{itemize}

\begin{figure*}[t]
    \centering
    \begin{subfigure}[b]{0.8\textwidth}
        \centering
        \includegraphics[width=\textwidth]{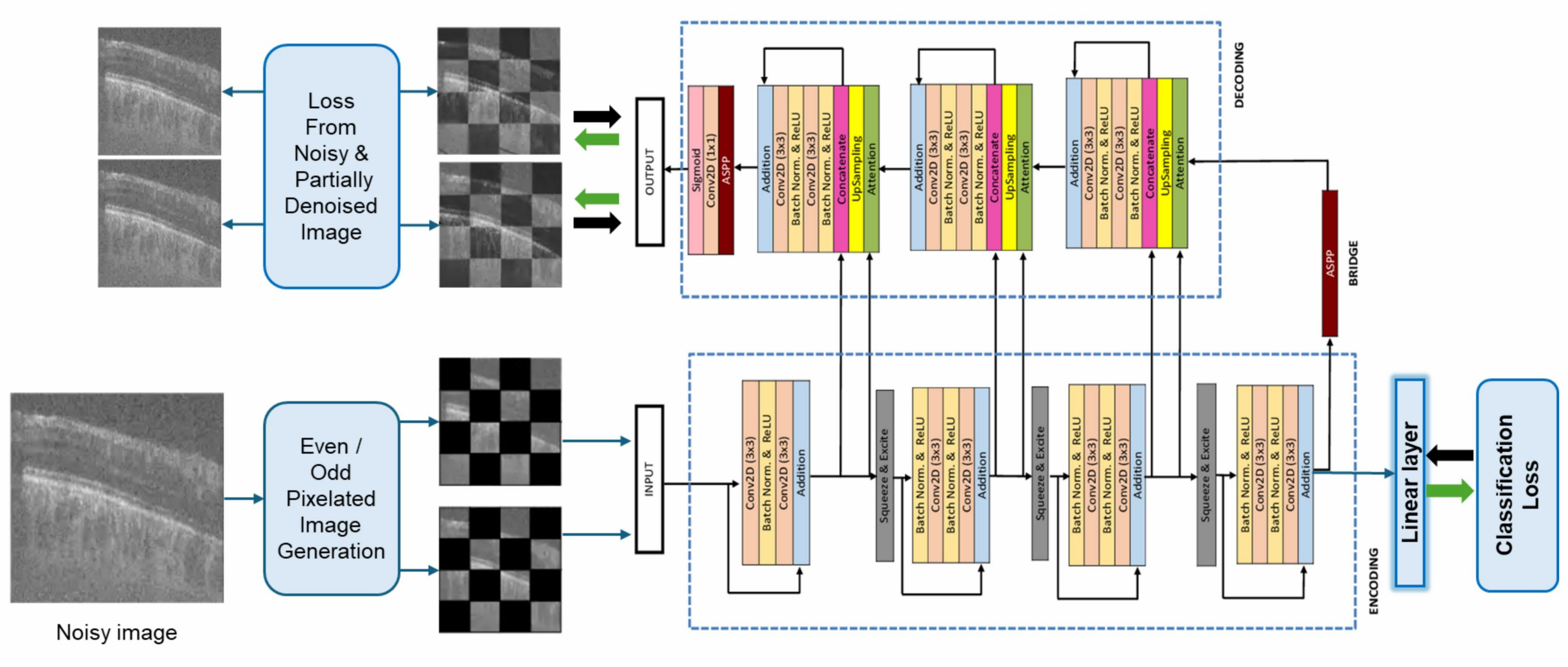}
        \caption{Model architecture for training}
        \label{fig:checkerboard_train}
    \end{subfigure}
    \vspace{1em}  
    \begin{subfigure}[b]{0.8\textwidth}
        \centering
        \includegraphics[width=\textwidth]{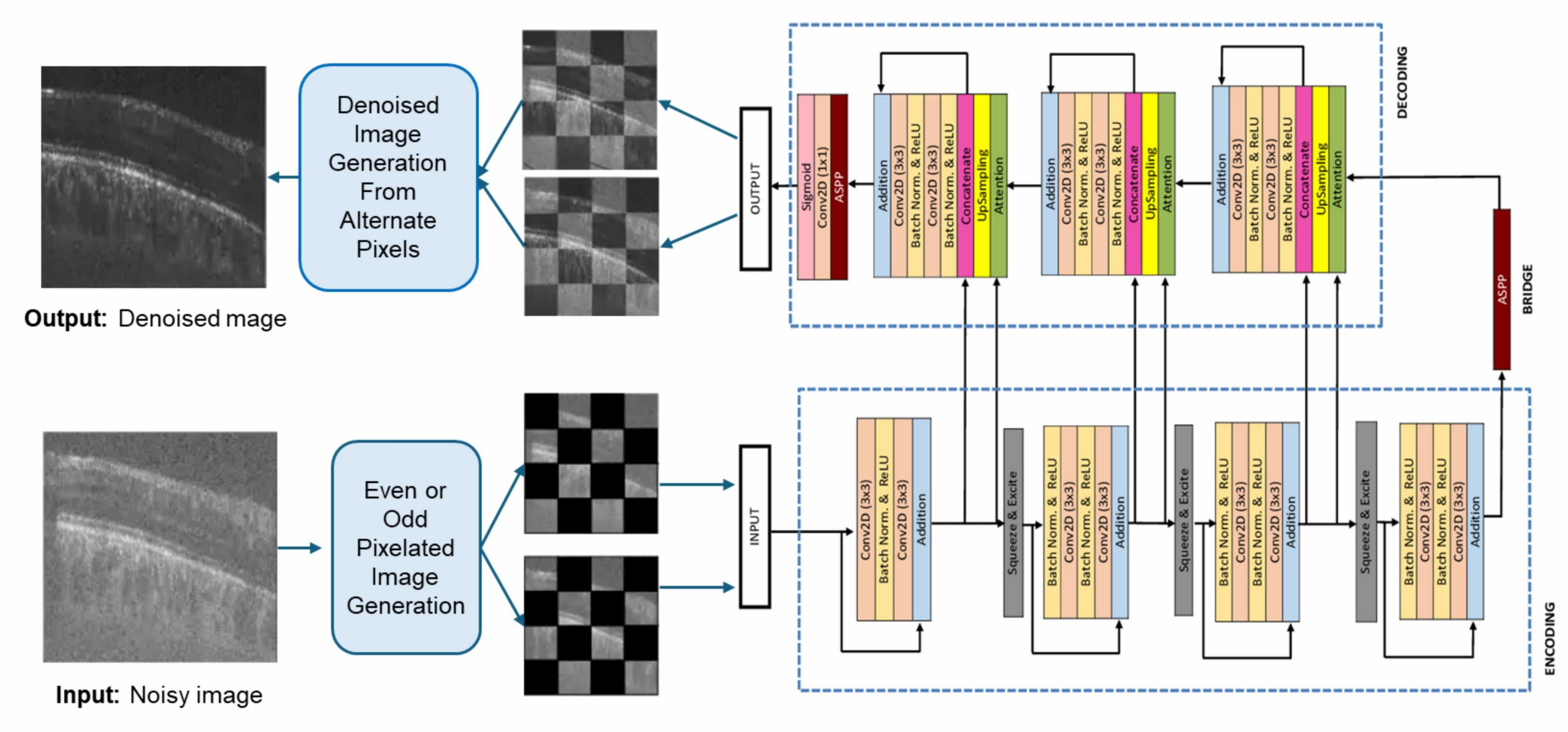}
        \caption{Model architecture for inference}
        \label{fig:checkerboard_infer}
    \end{subfigure}
    \caption{Checkerboard model architecture}
    \label{fig:checkerboard_models}
\end{figure*}

\section{Proposed Methodology for OCT Image Denoising and Classification}
In Figure \ref{fig:checkerboard_models}, we show the architecture of the proposed checkerboard model during training and inference. The checkerboard blind spotting technique is first applied to create a pair of blind-spotted images, one with odd positions and the other with even positions. During the training phase, two models are trained -- one for predicting odd blind-spotted pixels from even pixels, and the other for predicting even blind-spotted pixels from odd pixels. Finally, during inference, odd and even pixel predictions from the two models are fused together, to create the final denoised image. The following section provides a detailed discussion of this architecture. \\

\begin{figure}[t]
    \centering
    \begin{minipage}{0.6\textwidth} 
        \centering
        \begin{subfigure}[t]{0.33\textwidth} 
            \centering
            \includegraphics[width=\linewidth]{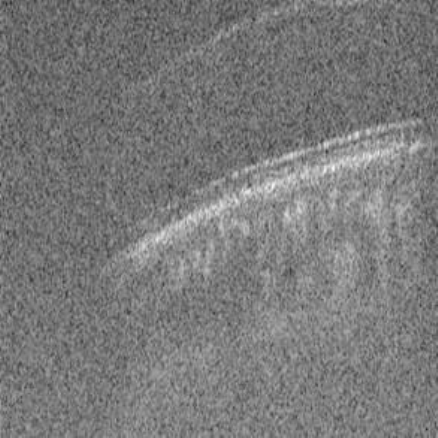}
            \caption{Original noisy OCT image}
            \label{fig:noisy_oct}
        \end{subfigure}\hfill
        \begin{subfigure}[t]{0.33\textwidth} 
            \centering
            \includegraphics[width=\linewidth]{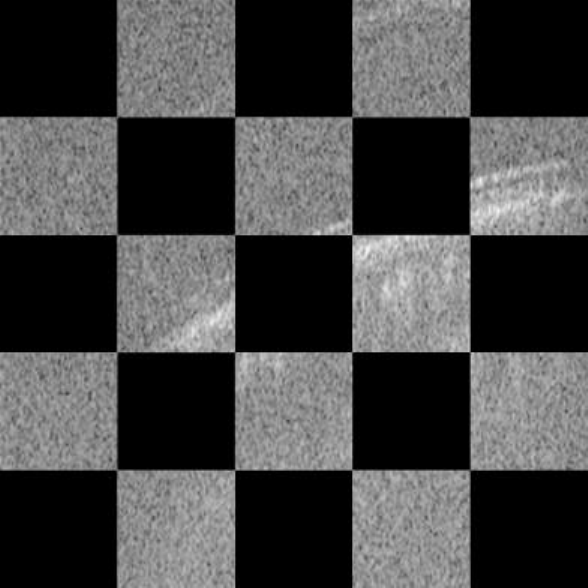}
            \caption{Odd-positioned pixels removed}
            \label{fig:odd_blind}
        \end{subfigure}\hfill
        \begin{subfigure}[t]{0.33\textwidth} 
            \centering
            \includegraphics[width=\linewidth]{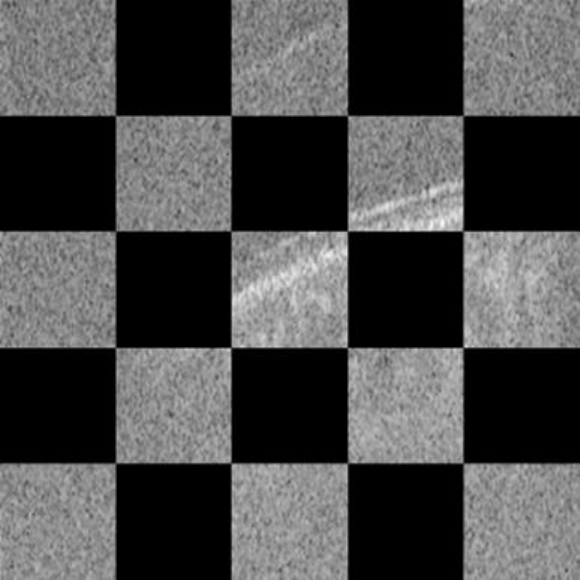}
            \caption{Even-positioned pixels removed}
            \label{fig:even_blind}
        \end{subfigure}
    \end{minipage}
    \caption{Checkerboard blind-spot creation}
    \label{fig:blindspot_making}
\end{figure}

\noindent\textbf{Checkerboard Blind-Spotting.} A noisy image is first blind spotted at even-positioned pixels and another one is blind spotted at odd-positioned pixels. In Figure \ref{fig:blindspot_making}, we show how the even and odd-positioned pixels are removed to create two blind-spotted versions of a noisy image\footnotemark[1]. \footnotetext[1]{This creates a checkerboard-like pattern at the pixel level; hence the name of our proposed approach.} By setting the blind spots in an alternating checkerboard arrangement, each blind spot pixel is surrounded by a grid of non-blind pixels. This structure in our methodology should ensure that every blind pixel has multiple neighboring clean pixels from all directions, allowing the model to utilize richer contextual information when making predictions. In contrast, methods similar to the Noise2Void~\cite{krull2019noise2void} or P2NV~\cite{krull2020probabilistic} use a local receptive field for prediction of a blind spotted pixel. This allows the model to use only the locally available context of the image. As the checkerboard blind-potting keeps pixels from all across the image, it enables the model to learn from a wider context, helping the model to retain important clinical details. 

Additionally, we note that multiple blind pixels can be predicted simultaneously across the image in a checkerboard pattern, since each blind pixel is surrounded by non-blind neighbors. This allows for parallel processing, which enables faster training compared to the conventional single-pixel masking approach. We argue that the proposed blind spotting approach contributes towards a more computationally efficient model training, which has the additional benefit of utilizing context from all across the image instead of local receptive field. We empirically verify this claim with our experimental results. \\

\noindent\textbf{Loss Function and Model Training. }As mentioned before, we utilize two losses to train our model. A Cross Entropy Loss is used that penalizes inaccurate classification labels of the image, and an RMS loss is used that penalizes the removal of clinical details. Overall, the loss function is given by: 
\[
	J = w_r \cdot \mathrm{RMSLoss} + w_c \cdot \mathrm{CrossEntropyLoss},
\]
where $w_r$ and $w_c$ are weights assigned for the RMSLoss and CrossEntropyLoss, respectively. In our experiments, we have used $(w_r, w_c) = (1, 0.2)$, which are chosen based on our hyper-parameter search. The RMSLoss is computed by comparing the predicted denoised image against the original noisy image. To calculate the CrossEntropyLoss, a linear fully connected layer is appended to the encoder block of the ResUNet++ architecture as shown in Figure \ref{fig:checkerboard_train}. As the encoder block extracts key features from the image, the linear layer's output is compared with the ground truth class of the noisy image to compute the CrossEntropyLoss. We note that our classification-aware design makes the proposed approach particularly effective in enhancing image quality without compromising diagnostic details, even when trained with limited medical image datasets.

With this loss function, we propose to train two models. The odd pixel predictor model is trained for even position blind-spotted images -- this model learns to predict odd-positioned pixels. The even pixel predictor model is trained on odd-position blind-spotted images, which learns to predict even-positioned pixels. \\

\noindent\textbf{Inference. }In Figure \ref{fig:checkerboard_infer}, we show the model architecture during inference. The noisy image is blind spotted at even and odd positioned pixels -- creating two copies. The odd and even pixel predictor models are used to predict the odd pixels and the even pixels. Their output is assembled into one image to produce the final denoised image. \\

\begin{figure}[t]
    \centering
    \includegraphics[width=0.7\columnwidth]{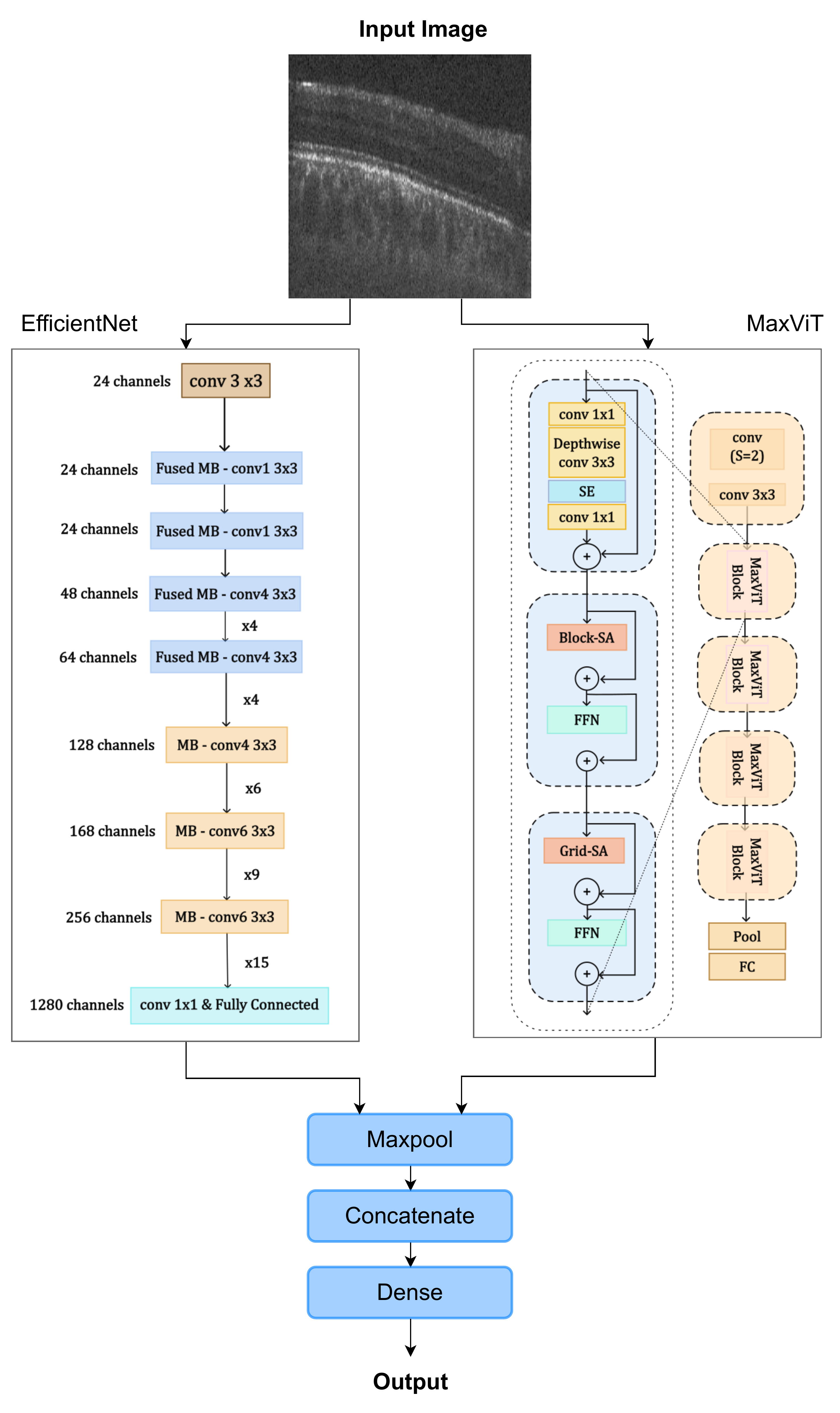}
    \caption{Mutual learning network for classification}
    \label{fig:mutual_learning}
\end{figure}

\noindent\textbf{Methodology for Classification. }As mentioned before, the effectiveness of the denoised OCT images in classification task is the key focus of our current work. For classification of the denoised images, we used a mutual learning technique~\cite{zhang2022semisupervised} to enhance overall classification accuracy. To this end, we note that CNN and Vision Transformers (ViT) are the two popular and effective classification network types. While their individual performance can be quite good, it can be even enhanced using mutual learning technique. More specifically, two models are trained together and hence can work together in mutual learning, to combine their strengths and produce better results. In this work, we chose a CNN based model and a ViT based model. The rationale behind our choice is two fold: CNNs are good for extracting local features and transformers were made for better extraction of global features. Hence, we argue that combining these two architectures allows our model to produce denoised OCT images that retain clinical features. As shown in Figure \ref{fig:mutual_learning}, the last layers of the EfficientNet and MaxViT has been connected to a dense layer after max-pooling, thus connecting the two models for mutual learning.

\begin{figure*}[t]
    \centering
    \begin{subfigure}[t]{0.18\textwidth}
        \centering
        \includegraphics[width=\linewidth]{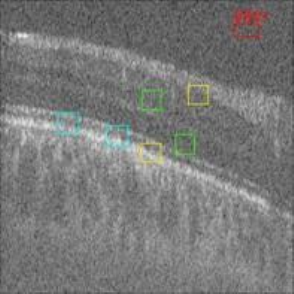}
        \caption{Noisy image with Undesired ROI (UROI) and Desired ROI (DROI) annotated. Green - DROI for CNR, MSR; blue - DROI for TP; yellow - DROI for EP.}
        \label{fig:result_noisy}
    \end{subfigure}
    \begin{subfigure}[t]{0.18\textwidth}
        \centering
        \includegraphics[width=\linewidth]{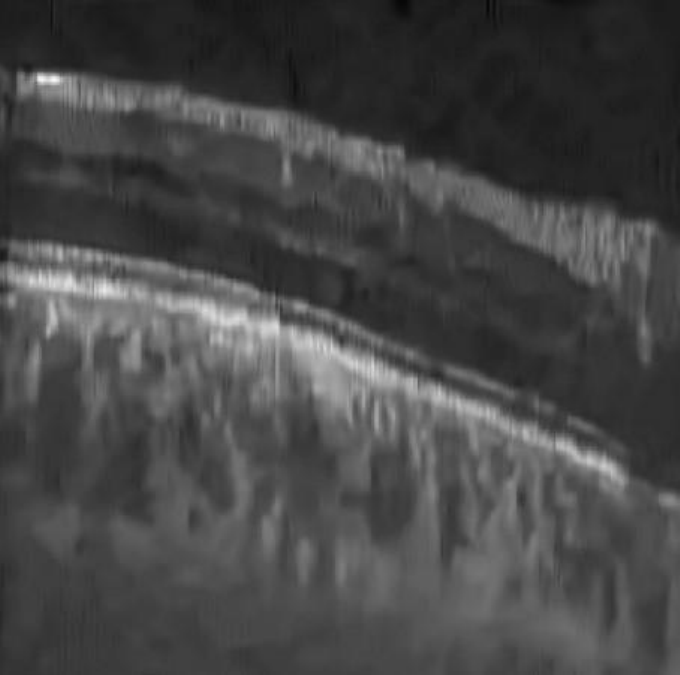}
        \caption{Denoised image using BM3D denoiser}
        \label{fig:result_bm3d}
    \end{subfigure}
    \begin{subfigure}[t]{0.18\textwidth}
        \centering
        \includegraphics[width=\linewidth]{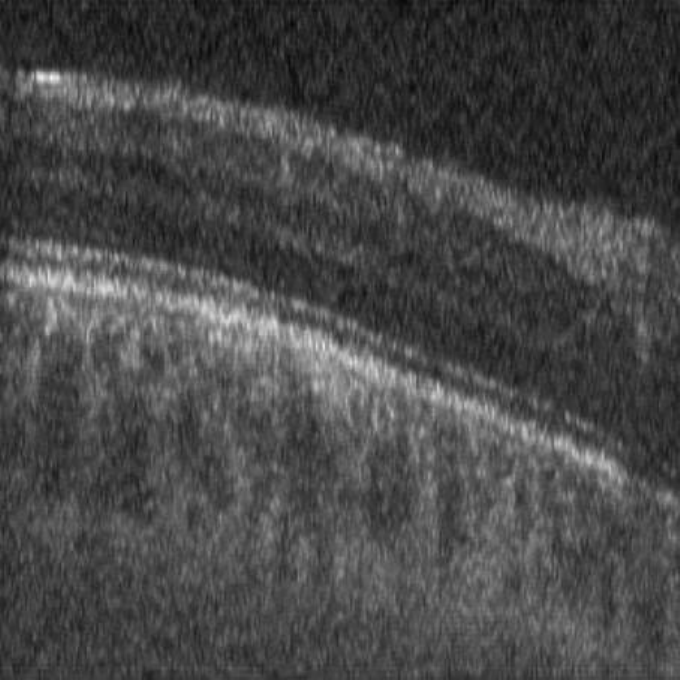}
        \caption{Denoised image using PN2V}
        \label{fig:result_p2nv}
    \end{subfigure}
    \begin{subfigure}[t]{0.18\textwidth}
        \centering
        \includegraphics[width=\linewidth]{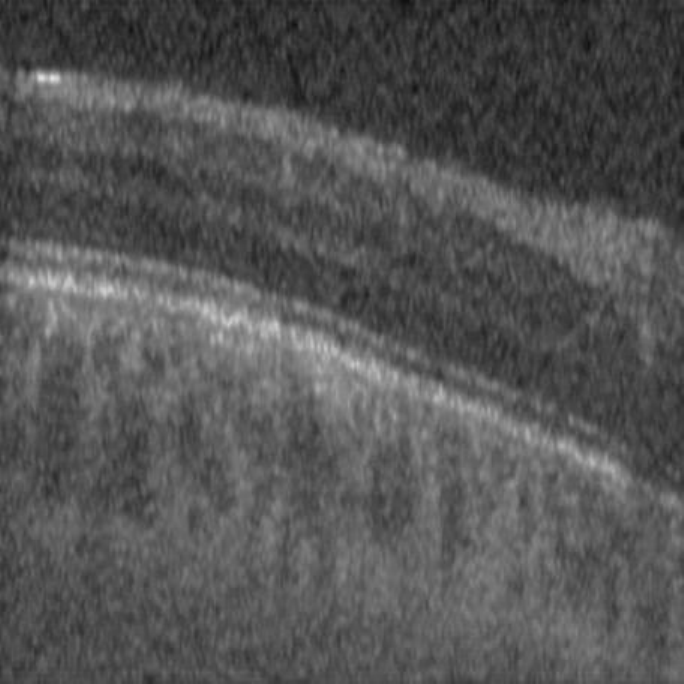}
        \caption{Denoised with n2v denoise}
        \label{fig:result_n2v}
    \end{subfigure}
    \begin{subfigure}[t]{0.18\textwidth}
        \centering
        \includegraphics[width=\linewidth]{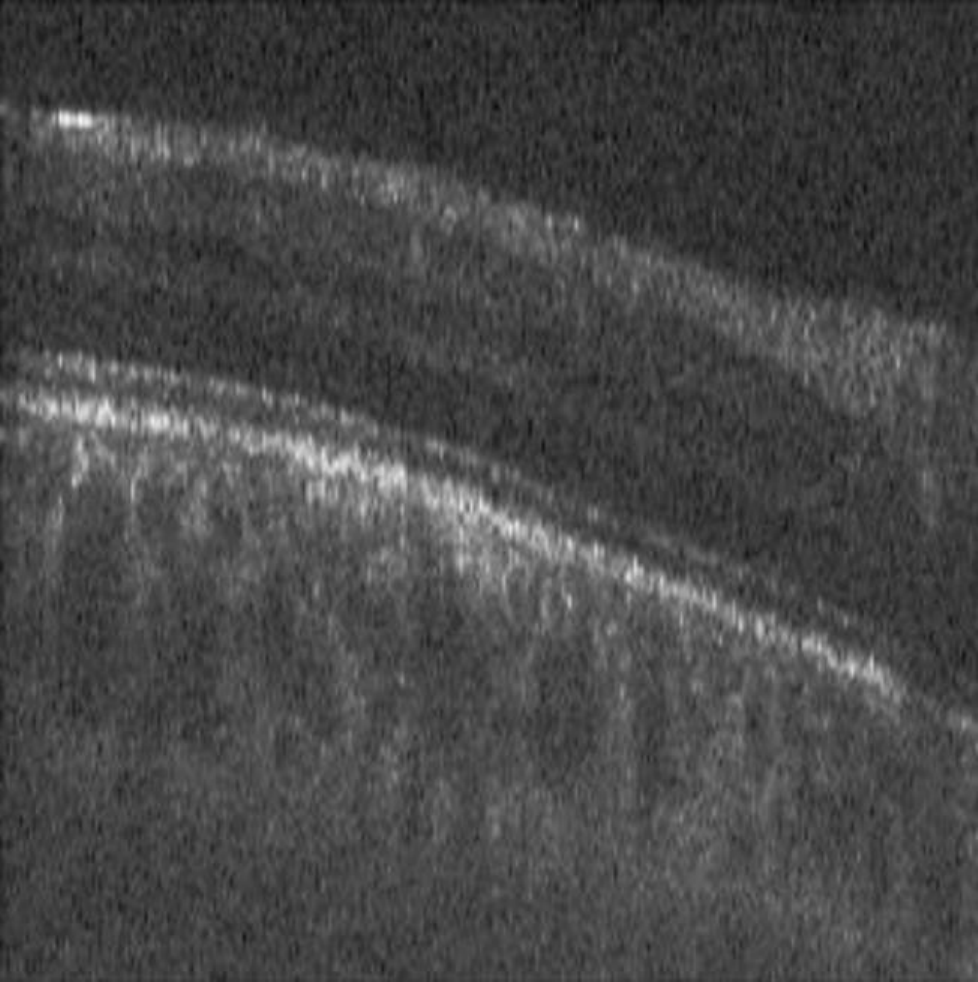}
        \caption{Denoised with checkerboard denoise (\textbf{proposed method})}
        \label{fig:result_checkerboard}
    \end{subfigure}
    \caption{Visual comparison of denoising results among existing models and our proposed model.}
    \label{fig:results}
\end{figure*}

\section{Experimental Results}
In this section, we demonstrate the performance of our proposed model on the VIP Cup 2024 training dataset \cite{icassp_vipcup2024}\cite{icassp_vipcup_dataset}. Unlike most medical image denoising datasets, this particular dataset contains classification information of images. This has enabled us to investigate the novel method of classification-aware denoising. The dataset includes 124 volume OCT data from 124 subjects. Each volume contains about 30 to 300 B-scans of size $300 \times 300$. The OCT data have been captured from different subjects and categorized into three groups: i) Healthy (class 0), ii) Diabetic with DME (class 1), and iii) Non-diabetic patients with other ocular diseases (class 2). All of our experiments were performed on one Tesla P100 GPU using the Adam optimizer along with multi-step learning rate adaptation. \\

\noindent\textbf{Performance indices for denoising. }We utilize four indices for evaluating the performance of our proposed model. In the following, we first briefly describe these indices and then present our results.

\begin{itemize}
    \item \textbf{Contrast-to-Noise Ratio (CNR)}~\cite{geissler2007contrast}: It is defined as $CNR = 10\log{\frac{\left|\mu_f - \mu_b\right|}{\sqrt{0.5\left(\sigma_f^2 + \sigma_b^2\right)}}}$, where $\sigma_b$ and $\mu_b$ denote the standard deviation and the mean of the background region, while $\mu_f$ and $\sigma_f$ represent the corresponding parameters for the foreground regions. The ROIs (Regions of Interest) for calculating CNR are chosen between different layers to show how the contrast changes.

    \item \textbf{Mean-to-Standard Deviation Ratio (MSR)}~\cite{wang2018adaptive}: It is defined as $MSR = \frac{\mu_f}{\sigma_f}$, where $\mu_f$ and $\sigma_f$ are the mean and the standard deviation of the foreground regions (these regions encompass the retinal layers).

    \item \textbf{Texture Preservation (TP)}~\cite{amini2017optical}: It is defined as $TP = \frac{\sigma_m^2}{\sigma_m^{\prime^2}} \sqrt{\frac{\mu_{den}}{\mu_{in}}}$, where $\sigma_m$ and $\sigma_m^\prime$ denote the standard deviation of the de-noised and noisy images in the $m$-th ROI. $\mu_{in}$ and $\mu_{den}$ represent the mean values of the noisy and de-noised images, respectively. The ROIs for calculating TP should encompass the intra-layer regions.

    \item \textbf{Edge Preservation (EP)}~\cite{amini2017optical}: It is defined as
    \begin{equation*}
    EP = \frac{\Gamma\left(\Delta I_m^\prime - \overline{\Delta I_m^\prime}, \Delta I_m - \overline{\Delta I_m}\right)}{\sqrt{\Gamma\left(\Delta I_m^\prime - \overline{\Delta I_m^\prime}, \Delta I_m^\prime - \overline{\Delta I_m^\prime}\right) \cdot \Gamma\left(\Delta I_m - \overline{\Delta I_m}, \Delta I_m - \overline{\Delta I_m}\right)}},
    \end{equation*}
    where $I_m$ and $I_m^\prime$ are matrices containing the de-noised and noisy image regions, respectively, in the $m$-th ROI. $\Delta$ represents the Laplacian operator, and $\Gamma$ measures the correlation between images. For the calculation of the EP metric, the relevant ROIs should encompass edge regions.
\end{itemize}

\noindent\textbf{Comparison of performance: denoising. }

\begin{table*}[t]
\caption{Comparison of performance: denoising}\label{tab:denoising_results}%
\small
\begin{tabular}{@{}p{2cm}ccccc@{}}
\toprule
Method & CNR & CNR (dB) & MSR & TP & EP \\
\midrule
Original Noisy Image & $0.6660 \pm 0.17$ & $-1.77$ & $7.62 \pm 0.66$ & $1.00 \pm 0.0$ & $1.00 \pm 0.0$ \\
BM3D~\cite{bm3d} & $1.75 \pm 0.25$ & $2.43$ & $6.96 \pm 0.89$ & $0.71 \pm 0.15$ & $0.47 \pm 0.08$ \\
PN2V~\cite{pn2v} & $2.49 \pm 0.11$ & $3.96$ & $7.06 \pm 0.47$ & $1.41 \pm 0.08$ & $\boldsymbol{0.90 \pm 0.08}$ \\
Noise2Void~\cite{n2v} & $2.30 \pm 0.15$ & $3.62$ & $7.23 \pm 0.50$ & $1.10 \pm 0.11$ & $0.61 \pm 0.09$ \\
\bfseries Checkerboard (our method) & $\boldsymbol{2.56 \pm 0.04}$ &  $\boldsymbol{4.08}$ &  $\boldsymbol{7.73 \pm 0.78}$ &  $\boldsymbol{1.87 \pm 0.03}$ &  $0.69 \pm 0.01$ \\
\botrule
\end{tabular}
\end{table*}

In Table~\ref{tab:denoising_results}, we show a comparison of performance among existing models and our proposed checkerboard model. As we can observe from the table, our proposed model outperforms the existing models in terms of all the performance indices defined above. We note that our proposed method shows significant superiority in terms of texture preservation. While the proposed method lags behind the PN2V approach~\cite{pn2v} in edge preservation, it achieves superior overall performance.

In Figure \ref{fig:results}, we present a visual comparison among these different models for a noisy image. The chosen ROI has been marked with color-coded rectangles. The undesired ROI (UROI) has been chosen in the background noisy portion, while the desired ROI (DROI) has been chosen between different retinal layers of the OCT image. As seen in Figure \ref{fig:result_bm3d} the BM3D method clearly smoothens out the denoised image. This aligns with the data in Table \ref{tab:denoising_results} showing poor texture preservation(TP) and edge preservation(EP) metrics for the BM3D metric. Visually, the PN2V method in Figure \ref{fig:result_p2nv} and our method in Figure \ref{fig:result_checkerboard} appear the clearest, whereas the N2V method in Figure \ref{fig:result_n2v} looks noticeably noisier. This also aligns with the experimental data in Table \ref{tab:denoising_results} which suggests a close performance metric between PN2V and our method. However, as we will see in the next section, while our method achieves comparable denoising performance with PN2V, it excels in classification from denoised images thanks to its classification-aware denoising approach.\\

\noindent\textbf{Comparison of performance: classification. }We now demonstrate the classification performance on the denoised images generated by our proposed model. As mentioned before, we employed the mutual learning mechanism for strong classification performance. To select the optimal models for the mutual learning network, we tested various CNN-based and ViT-based architectures. Among the CNN models, EfficientNet V2-M provided the best results, while MaxVit-L performed best among the ViT models. The experimental results (performed on the images denoised with our proposed checkerboard model) are shown in Table~\ref{tab:classification_results}. The results indicate that using the mutual learning technique improves image-wise accuracy by 0.8\%, while providing the same subject-wise accuracy. 

Additionally, to highlight the effectiveness of our proposed denoising method in classification awareness over other denoising methods, we performed classification after denoising by the several existing denoising methods, and the results are listed in Table ~\ref{tab:classification_res_for_denoising_methods}. We can observe that denoised image produced by our proposed method provides the best classification result. Finally, we note from Table~\ref{tab:denoising_results} that the proposed checkerboard method achieves a better average CNR by approximately 2.8 \%, whereas a better subject-wise accuracy by 5.2\% and a better image-wise accuracy by 4.4\%. This clearly underlines the superiority of our proposed method in classification awareness, which eventually translates into better diagnosis results with the denoised images.

\begin{table}[t]
\caption{Comparison of performance: classification (with and without Mutual Learning)}\label{tab:classification_results}
\small
\begin{tabular}{@{}p{3.5cm}cccc@{}}
\toprule
Model & Subj.-wise Acc. & Img.-wise Acc. & Prec. & Recall \\ 
\midrule
EfficientNet V2-M~\cite{tan2021efficientnetv2} & \textbf{95.0}\% & 84.8\% & 84.8\% & 84.5\% \\ 
MaxVit-L~\cite{tu2022maxvit} & 85.0\% & 79.8\% & 79.8\% & 79.5\% \\ 
\textbf{Proposed training (Mutual Learning with EfficientNet V2-M and MaxVit-L)} & \textbf{95.0}\% & \textbf{85.6}\% & \textbf{85.6}\% & \textbf{85.5}\% \\ 
\botrule
\end{tabular}
\end{table}

\begin{table}[t]
\caption{Comparison of performance: classification (after different denoising methods are applied)}\label{tab:classification_res_for_denoising_methods}
\small
\begin{tabular}{@{}p{3cm}cccc@{}}
\toprule
Denoising Model & Subj-wise Accuracy & Img-wise Accuracy & Precision & Recall \\ 
\midrule
Original Noisy Image & 81.2\% & 73.3\% & 79.0\% & 75.0\% \\ 
BM3D~\cite{bm3d} & 79.5\% & 71.2\% & 77.5\% & 73.5\% \\ 
PN2V~\cite{pn2v} & 89.2\% & 81.2\% & 88.0\% & 85.0\% \\ 
Noise2Void~\cite{n2v} & 87.1\% & 79.3\% & 86.0\% & 83.0\% \\ 
\textbf{Checkerboard (our method)} & \textbf{95.0\%} & \textbf{85.6\%} & \textbf{93.0\%} & \textbf{91.0\%} \\ 
\botrule
\end{tabular}
\end{table}

\section{Conclusion}
In this work, we introduced a novel denoising method that addresses three crucial demands of medical image denoising -- not introducing new artifacts, not altering important clinical details, and trainability in self-supervised fashion from single noisy images. We empirically demonstrated that our method performed superbly compared to the existing denoising methods. We argue that our proposed network and training approach provides an additional benefit of adapting to any type of noise present in the dataset unlike existing statistical methods. In essence, our work encompasses a broader aspect of the medical image denoising and enhancement problem -- the proposed contextual checkerboard denoising approach can be implemented for other datasets/applications, providing state-of-the-art denoising while preserving the intricate details of the images. This inevitably results in better diagnosis through image denoising.

\section*{Declarations}
\begin{itemize}
    \item \textbf{Funding}:
    No funds, grants, or other support was received.
    
    \item \textbf{Conflict of interest}:
    The authors have no relevant financial or non-financial interests to disclose.
    
    \item \textbf{Data availability}: The dataset used is available on the URL: https://misp.mui.ac.ir/en/ieee-video-and-image-processing-cup. A mirror of the dataset can be found here: https://www.kaggle.com/datasets/touhid31416/vip-cup-2024-train-test-dataset. 
    
    \item \textbf{Ethics and Consent to Participate declarations}: not applicable

\end{itemize}
\noindent


\section*{Acknowledgment}
The authors would like to express their sincere gratitude towards the Department of Electrical and Electronic Engineering (EEE) of Bangladesh University of Engineering and Technology (BUET) for providing support for research. 

\bibliography{sn-bibliography}

\end{document}